\documentclass[
    ,final            
  ]
  {aipproc}

\layoutstyle{8x11double}

\newcommand{\bl}{\begin{itemize}}
\newcommand{\el}{\end{itemize}}
\newcommand{\be}{\[}
\newcommand{\ee}{\]}
\newcommand{\bq}{\[}
\newcommand{\eq}{\]}
\newcommand{\ba}{\begin{eqnarray*}}
\newcommand{\ea}{\end{eqnarray*}}
\newcommand{\bqa}{\begin{eqnarray*}}
\newcommand{\eqa}{\end{eqnarray*}}

\newcommand{\PhysRevD}[3]{Phys. Rev. {\bf D #1}, #2 (#3)}
\newcommand{\NuclPhysB}[3]{Nucl. Phys. {\bf B #1}, #2 (#3)}
\newcommand{\PhysLettB}[3]{Phys. Lett. {\bf B #1}, #2 (#3)}

\begin{document}

\title{Sudakov Expansions and Top Quark Physics at LHC}


\author{Matteo Beccaria}{
  address={Dipartimento di Fisica and INFN, Via Arnesano Ex Coll. Fiorini, 73100 Lecce, ITALY}
}

\begin{abstract}
We review some peculiar features of Sudakov expansions in the calculation of 
electroweak radiative corrections in the MSSM at high energy. We give specific examples
and consider in particular the process $bg\to tW$ of single top quark production relevant 
for the top quark physics programme at LHC.
\end{abstract}

\maketitle


\section{Introduction}

The analysis of physical processes in the MSSM is complicated by 
the large number of parameters of the model. Also, radiative correction
typically involve a large subset of them making apparently quite difficult
to isolate specific physical effects. 
However, there is a way out to simplify the situation at the price
of some controlled approximation. Here we shall be concerned with the 
electroweak sector of the MSSM. The planned future collider
experiments (LHC, ILC) will be characterized by large invariant masses and 
at very high energy it is known that radiative corrections simplify and 
become smooth, at least beyond production thresholds. In this regime they can be 
described by rather simple asymptotic expansions. These are basically
power series in the logarithm of the center of mass energy or typical large
invariant mass.
This logarithmic approximation is called a Sudakov expansion and 
requires light SUSY scenarios with not too heavy sparticles in order to be accurate~\cite{LSE}.
Apart from this basic requirement, it compactly expresses the relevant  
radiative corrections with a very minimal set of parameters~\cite{TanBeta}. 
The deep reason for this important feature is that at high energy 
soft SUSY breaking operators are suppressed for dimensional reasons
and disappear at the leading orders in the expansion.

The origin of large logarithms in the physical amplitudes is nowadays 
quite clear~\cite{Beccaria:2003yn}.
There are two different kinds of contributions coming respectively from 
short and long distance and therefore of ultraviolet or infrared nature.
The UV logarithms are governed by standard Renormalization Group equations.
They typically arise in gauge boson self energy corrections or in 
essentially gaugeless sectors of the MSSM, like the important Yukawa one.
The IR logarithms are manifestations of mass singularities. They can arise in diagrams with
exchange of gauge bosons. In the calculation of exclusive processes the mass of
massive gauge bosons plays the role of an infrared regulator and gives rise to 
energy growing contributions of logarithmic type. For instance, at one loop
the leading terms are proportional to $\alpha\log^2(s/M^2)$ where $\alpha = e^2/4\pi$, 
$\sqrt{s}$ is the center of mass energy and $M$ is a process dependent scale.
The unusual counting of two logs per loop is a typical remnant of the IR origin of these contributions.

In the next Section, we shall discuss briefly the classification of the Sudakov
logarithms. This will provide the necessary tools to discuss the corrections to 
single top quark production processes at LHC, as an interesting application.

\section{Some classification of Sudakov Logarithms}

The electroweak Sudakov logarithms in a $2\to 2$ process at one loop can be classified according to the 
following three main categories. We closely follow the notation of~\cite{Beccaria:2003yn}
and refer to this paper for minor or notational details concerning the next paragraphs.

\paragraph{a) UV Logarithms from gauge boson self energies}

If $F^{Born}$ is the amplitude under study, at one loop we get from coupling constants renormalization,
the following correction term 
\be
F^{RG} = -\frac{1}{4\pi^2}\left( g^4 \beta\ \frac{\partial F^{Born}}{\partial g^2} + 
g^{'\ 4} \beta'\ \frac{\partial F^{Born}}{\partial g^{'\ 2}} \right)\ \log\frac{s}{\mu^2}
\ee
and of course $\beta, \beta'$ depend on the model (SM, MSSM, Split SUSY, etc.)
 
\paragraph{b) Universal Logarithms}

These are terms independent on the scattering angles. They can be computed 
by analyzing the various external lines associated to initial of final states
one after the other. Each line gets a correction factor that we now discuss.
The general form of the correction is ($p$ = scalar, fermion or vector)
\be
\frac{\alpha}{\pi}\ c_p^{\rm gauge}\ \left(n\log \frac{s}{M_V^2}-\log^2\frac{s}{M_V^2}\right) + \frac{\alpha}{\pi} c^{\rm Yukawa}_p\ \log\frac{s}{M^{'\ 2}}
\ee
where $V = \gamma, Z, W^\pm$. The linear logarithm scale is a choice to be fixed at NNLO.

In each model and for each kind of external line (initial of final) we can list the various coefficients
($n$, $c_p$, $c^{\rm Yukawa}$) and write immediately the universal corrections at one loop and next to leading 
logarithmic order.

\paragraph{c) Angular dependent logarithms}

The final type of logarithms is called angular dependent since depends on the 
scattering angle. These are terms of the form 
\be
\frac{\alpha}{\pi}\log\frac{s}{M^2}\log\frac{1\pm\cos\vartheta}{2},
\ee
where $\vartheta$ is the scattering angle. These contributions arise from Standard Model box diagrams
and do not receive SUSY additional terms.

\subsection{An example: correction to  fermion or sfermion external lines}

To give an explicit example of the above mentioned corrections, we now consider the specific case of 
external lines associated to a fermion (lepton or quark) or a sfermion (slepton or squark). The universal logarithms are
\be
\frac{\alpha}{\pi}\ c^{\rm gauge}\ \left(n\log \frac{s}{M_V^2}-\log^2\frac{s}{M_V^2}\right) + \frac{\alpha}{\pi} c^{\rm Yukawa}\ \log\frac{s}{m_t^2}
\ee
where $m_t$ is the top quark mass. The gauge part reflects the SU(2) $\times$ U(1) structure. With standard notation, it reads
\ba
c_f^{\rm gauge} &=& \frac{1}{8}\left[\frac{I_f(I_f+1)}{s_W^2} + \frac{Y_f^2}{4c_W^2}\right] \\
Y_f &=& 2(Q_f-I_f^3) 
\ea
with $n = 3$ in the Standard Model and $n = 2$ in the MSSM.
The Yukawa part is present only for heavy quarks top and bottom and reads
(the upper line refers to the Standard Model, the lower to the MSSM)
\be
c_{b_L} = c_{t_L} = \left\{ \begin{array}{l} 
\displaystyle -\frac{1}{32 s_W^2}\left(\frac{m_t^2}{M_W^2} + \frac{m_b^2}{M_W^2}\right)  \\ \\
\displaystyle -\frac{1}{16 s_W^2}\left(\frac{m_t^2}{M_W^2}\frac{1}{\sin^2\beta} + \frac{m_b^2}{M_W^2}\frac{1}{\cos^2\beta} \right) 
\end{array}\right.
\ee

\be
c_{b_R} = \left\{ \begin{array}{l} 
\displaystyle -\frac{1}{16 s_W^2}\ \frac{m_b^2}{M_W^2}  \\ \\
\displaystyle -\frac{1}{8 s_W^2}\ \frac{m_b^2}{M_W^2}\frac{1}{\cos^2\beta}
\end{array}\right.
c_{t_R} = \left\{ \begin{array}{l} 
\displaystyle -\frac{1}{16 s_W^2}\ \frac{m_t^2}{M_W^2}  \\ \\ 
\displaystyle -\frac{1}{8 s_W^2}\ \frac{m_t^2}{M_W^2}\frac{1}{\sin^2\beta}
\end{array}\right.
\ee
The angle $\beta$ is the common MSSM parameter defined at tree level by $\tan\beta = v_2/v_1$
where $v_{1,2}$ are the vevs of the two Higgs doublets.

For sfermions (sleptons or squarks) the gauge part does not change. Also the Yukawa 
terms are identical (matching chiralities). This is a non trivial consequence of supersymmetry and of the 
general fact that breaking tends to be suppressed at high energy.

\section{Single Top production at LHC: Sudakov corrections}

Single top quark production at LHC is a very interesting topic for the search of 
New Physics effects via radiative corrections. It is also remarkable due to its 
special feature of being a direct measure of $|V_{tb}|^2$. At leading order there are three 
basic processes that must be considered, i.e. at parton level, the processes
$bu\to td$, $u\overline{d}\to t\overline{b}$ and $bg\to tW$. They have quite different features as discussed in details 
in~\cite{Beneke:2000hk}. Application of the Sudakov technique to the three processes have been discussed
in~\cite{Beccaria:2004xk}. Here we discuss in some details the associated $Wt$ production 
$bg\to tW^-$ as a complete application of the previous general discussion.

\begin{figure}
  \includegraphics[height=2cm]{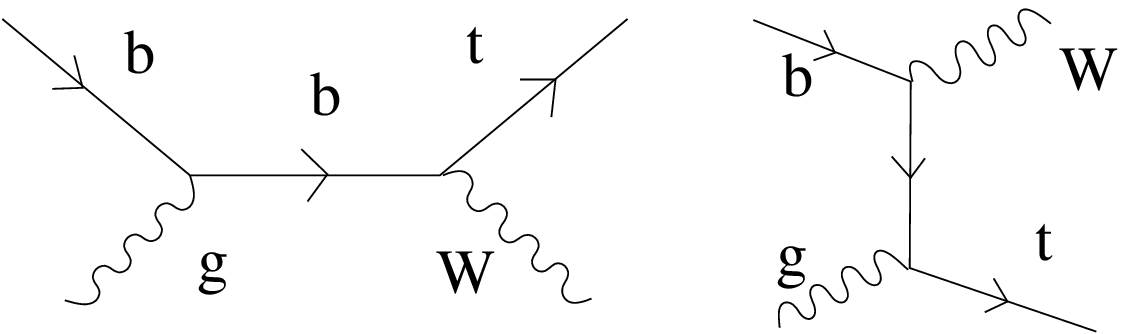}
  \caption{$bg\to tW$ at tree level.}
  \label{fig:tree}
\end{figure}

The two tree level diagrams to be considered are shown in Fig.~(\ref{fig:tree}).
In the first ($a$) a bottom quark is exchanged in the $s$ channel; In the second ($b$) 
a top quark is exchanged in the $u=(p_b-p_W)^2$ channel. To simplify the discussion let us
neglect all masses with the exception of $m_t$. We neglect the ratio $m_t/\sqrt{s}$, but 
obviously keep $m_t/M_W$. 
Let us denote the helicity amplitude for the process as $F_{\alpha\alpha'\beta\beta'}$ where 
the four subscripts denote the helicities of the four external states $b$, $g$, $t$, and $W$.
In our approximations we remain with the two non suppressed helicity amplitudes
\bq
F^{\rm Born~a+b}_{----}\to g_{WL}\ g_s({\lambda^l\over2})
{2\over \cos{\vartheta\over2}}
\eq
\bq
F^{\rm Born~a+b}_{-+-+}\to g_{WL}\ g_s({\lambda^l\over2})
{2 \cos{\vartheta\over2}}
\eq
\bq
F^{\rm Born~a+b}_{-++0}\to g_{WL}\ g_s({\lambda^l\over2})
\sqrt{2}\ {m_t\over M_W}
\cos{\vartheta\over2}\ {1-\cos\vartheta\over1+\cos\vartheta}
\label{fbornl}\eq
The differential cross section is simply ($s, t, u$ are standard Mandelstam variables)
\bq
{d\sigma^{\rm Born}\over d\cos\vartheta}\to
-~{\pi \alpha\alpha_s\over24s^2_Wus^2}\left[s^2+u^2+{m^2_tt^2\over2M^2_W}\right]
\eq
We can now write in a very simple way the Sudakov electroweak corrections to these two
non suppressed amplitudes.

The universal component for producing a transverse $W$ is
\bqa
\frac{F^{\rm Univ}_{-,\mu,-,\mu}}{F^{\rm Born}_{-,\mu,-,\mu}} &=&
~{1\over2}~(~c^{\rm ew}(b\bar b)_{L}
+c^{\rm ew}(t\bar t)_{L}~)~+c^{\rm ew}(W_T)
\eqa
where $c^{\rm ew}(f\bar f)$ is the sum of the gauge and Yukawa correction associated to 
a $f$ fermion line. We have also introduced the new coefficient
\bq
c^{\rm ew}(W_T)={\alpha\over4\pi s^2_W}\left(-\log^2\frac{s}{M_W^2}\right)
\eq
In the production of longitudinal $W^-_0$ we have instead 
\bqa
\frac{F^{\rm Univ}_{-,+,+,0}}{F^{\rm Born}_{-,+,+,0}} &=&
~{1\over2}~(~c^{\rm ew}(b\bar b)_{L}
+c^{\rm ew}(t\bar t)_{R}~)~+c^{\rm ew}(W_0)
\label{funivl}
\eqa
where 
\bqa
c^{\rm ew}(W_0)&=&{\alpha\over\pi}({1+2c^2_W\over32s^2_Wc^2_W})
\left(n_G\log\frac{s}{M_W^2}-\log^2\frac{s}{M_W^2}\right)
\label{cewl}\eqa
and $n_G=4,0$ in the Standard Model or MSSM, respectively.

For reasons of space, we do not write the explicit angular dependent terms.
Finally there are SUSY QCD Sudakov logarithms from vertices
with gluino exchanges. They read
\bqa
F^{\rm Univ~SUSYQCD}_{-,\mu,-,\mu}&=&
F^{\rm Born}_{-,\mu,-,\mu}
\left(-\frac{\alpha_s}{3\pi}\log\frac{s}{M^2_{\rm SUSY}}\right)
\eqa
\bqa
F^{\rm Univ~SUSYQCD}_{-,+,+,0}&=&
F^{\rm Born}_{-,+,+,0}
\left(-\frac{\alpha_s}{3\pi}\log\frac{s}{M^2_{\rm SUSY}}\right)
\eqa
In this specific process there are no RG logarithms.

\section{Numerical Results}

The same set of corrections can be computed for the other three processes 
of single top production. The partonic cross sections are then converted in 
hadronic observables. In particular, one defines the distribution 
\bqa
\lefteqn{{d\sigma(PP\to t Y+...)\over ds} = } && \\ 
&& {1\over S}~\int^{\cos\overline\vartheta}_{-\cos\overline\vartheta}
d\cos\vartheta~\  ~\sum_{ij}~L_{ij}(\tau, \cos\vartheta)
{d\sigma_{ij\to  t Y}\over d\cos\vartheta}(s)~]
\eqa
where $\sqrt{S}$ is the  $pp$ energy, $L_{ij}$ is the luminosity from partons $i$, $j$
and the various kinematical variables are explained in details in~\cite{Beccaria:2004xk}.

\begin{figure}
  \includegraphics[height=5.5cm]{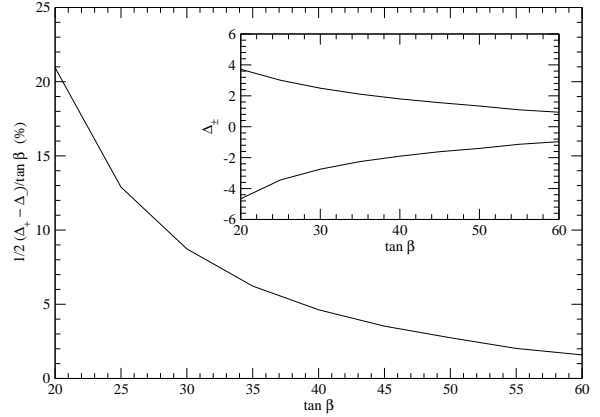}
  \caption{Bounds on $\tan\beta$ from the combination of the three single top production processes.}
  \label{fig:tanbeta}
\end{figure}

\begin{figure}
  \includegraphics[height=10cm]{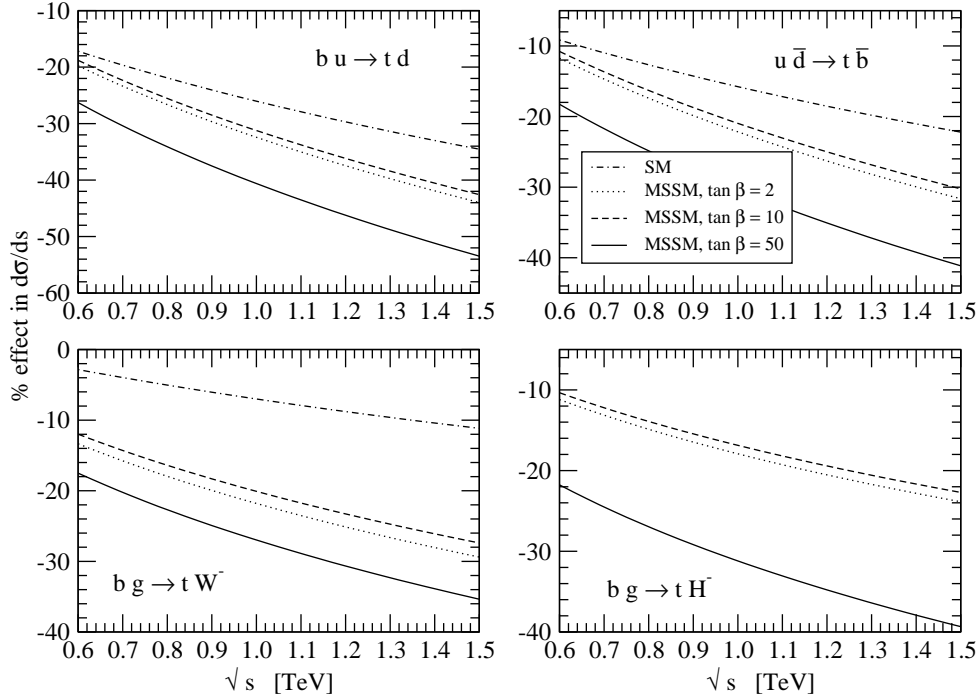}
  \caption{Effects in $d\sigma/ds$ for the three single top production processes.}
  \label{fig:effects}
\end{figure}

The results for the three single top production processes are summarized in Fig.~(\ref{fig:effects}).
Large effects can be obtained due to the Yukawa terms. These effects isolate the parameter 
$\tan\beta$ and can be used to fix bounds on it. To give an example of such a procedure, we 
have combined the effects in the  three production  processes assuming 
measurements in the range $\sqrt{s} = 500 - 1500$ GeV (with 20 GeV spacing) and an 
overall precision of 10\%. Fig.~(\ref{fig:tanbeta}) illustrates the results and
show the values of 
$\Delta_\pm$ which appear in the confidence region $(\tan\beta + \Delta_-, \tan\beta + \Delta_+)$ which is 
determined by a $\chi^2$ analysis of fake data simulated at a certain $\tan\beta$.

In practice, the simple Sudakov approximation 
displays interesting features and suggest that a full one loop calculation 
would be certainly worthwhile.

A similar analysis in the case of top - antitop pair production 
has also been completed and can be found in~\cite{Beccaria:2004sx}.

\end{document}